\documentclass[onecolumn]{pasj01}
\usepackage{comment}

\begin{document} 
\Received{2018/07/20}
\Accepted{2018/08/31}

\title{CO observations toward the isolated mid-infrared bubble {S44}: External triggering of {O-star} formation by a cloud-cloud collision}

\author{Mikito \textsc{Kohno}\altaffilmark{1}$^{*}$}
\altaffiltext{1}{Department of Physics, Nagoya University, Furo-cho, Chikusa-ku, Nagoya, Aichi 464-8601, Japan}
\altaffiltext{2}{Nobeyama Radio Observatory, National Astronomical Observatory of Japan (NAOJ), National Institutes of Natural Sciences (NINS), 462-2, Nobeyama, Minamimaki, Minamisaku, Nagano 384-1305, Japan}
\altaffiltext{3}{Institute for Advanced Research, Nagoya University, Furo-cho, Chikusa-ku, Nagoya 464-8601, Japan}
\email{mikito@a.phys.nagoya-u.ac.jp}
\email{k.tachihara@a.phys.nagoya-u.ac.jp}
\email{fujita.shinji@a.phys.nagoya-u.ac.jp}
\author{Kengo \textsc{Tachihara}\altaffilmark{1}$^{*}$}
\author{Shinji \textsc{Fujita}\altaffilmark{1}$^{*}$}
\author{Yusuke \textsc{Hattori}\altaffilmark{1}}
\author{Kazufumi \textsc{Torii}\altaffilmark{2}}
\author{Atsushi \textsc{Nishimura}\altaffilmark{1}}
\author{Misaki \textsc{Hanaoka}\altaffilmark{1}}
\author{Satoshi \textsc{Yoshiike}\altaffilmark{1}}
\author{Rei \textsc{Enokiya}\altaffilmark{1}}
\author{Keisuke \textsc{Hasegawa}\altaffilmark{1}}
\author{Akio \textsc{Ohama}\altaffilmark{1}}
\author{Hidetoshi \textsc{Sano}\altaffilmark{1,3}}
\author{Hiroaki \textsc{Yamamoto}\altaffilmark{1}}
\author{Yasuo \textsc{Fukui}\altaffilmark{1,3}}%



\maketitle

\begin{abstract}
{We {have performed} a multi-wavelength study of the mid-infrared bubble {S44} to investigate the origin of isolated high-mass star(s) and the star-formation {process} around the bubble {formed by the} H\,\emissiontype{II} region. In this paper, we report the results of new CO observations ($^{12}$CO, $^{13}$CO $J=$1--0, and $^{12}$CO $J=$3--2) toward the isolated bubble {S44} {using} the NANTEN2, Mopra, and ASTE radio telescopes.
We found two velocity components {in the direction of the bubble,} at $-84$ km s$^{-1}$ and $-79$ km s$^{-1}$. These two clouds are likely to be physically associated with the bubble{,both} because
 of the enhanced $^{12}$CO $J=$3--2/1--0 intensity ratio {from a} ring-like structure affected by ultraviolet radiation from embedded high-mass star(s) and {from the} morphological correspondence between the 8 $\mu$m emission and the CO distribution. 
{Assuming a single object, we estimate} the spectral type of the embedded star inside the bubble {to be} {O8.5-9 ($\sim 20 M_{\odot}$)} from the radio-continuum free-free emission. We {hypothesize} that the two clouds collided with each other {3 Myr} ago, {triggering} the formation of the isolated high-mass star {in} {S44}, as {also occurred} with RCW 120 and RCW 79. {We argue that this scenario can explain the origin of the isolated O-star inside the bubble. }}
\end{abstract}

\section{Introduction}

\subsection{{H\,\emissiontype{II} regions and {\it Spitzer} mid-infrared bubbles in the Milky Way}}
{H\,\emissiontype{II} regions are formed around high-mass stars, {which are} mainly {distributed} in the spiral {arms of} the Milky Way. They {ionize and} destroy {the} parent molecular clouds and {create cavity-like structures} around the exciting stars via stellar wind{s} and ultraviolet radiation (e.g., Whitworth 1979). These cavity-like structures in the interstellar medium (ISM) are often called “interstellar {bubbles}" (e.g., Castor et al. 1975b, Weaver et al. 1977).}

{The {\it Spitzer} mid-infrared bubbles were {identified} by Churchwell et al. (2006, 2007) {from} the Galactic Legacy Infrared Mid-Plane Survey Extraordinaire (GLIMPSE) data. They cataloged about 600 bubbles in {the} northern and southern hemisphere{s} in the Galactic plane {($|l|$$\leq$\timeform{65D}, $|b|$$\leq$ \timeform{1D}).}
The authors suggested that most bubbles are H\,\emissiontype{II} regions {that} contain embedded OB-type stars or star clusters.}

{Several formation scenarios have been discussed for the mid-infrared bubbles, including radiation-driven implosions (RDIs: Sandford et al 1982, Lefloch \& Lazareff 1994) and {the} collect-and-collapse process (C\&C: Elmegreen \& Lada 1977, Whitworth et al. 1994a). The first involves the compression of pre-existing clouds by the pressure of the ionized gas, while the second consists of sweeping-up of the diffuse ISM inside the wall of the expanding shell that is undergoing} gravitational collapse. These scenarios can explain {star formation at the edges of {bubbles}, which is triggered by the expanding {H\,\emissiontype{II} regions} (e.g., Deharveng et al. 2005, 2008, 2009, 2010, Zavagno et al. 2006, 2007). On the other hand, from their numerical simulations, Dale et al. (2015) have pointed out that it is difficult to distinguish between triggered and spontaneous star formation around H\,\emissiontype{II} regions solely from their observational morphologies.}



{Recently, Torii et al. (2015) carried out CO observations toward the mid-infrared bubble RCW 120. They {showed} that a cloud-cloud collision scenario can explain the {morphologies} of the bubbles and {{the} formation of} {ionizing O stars}, based on numerical simulation{s} of a head-on collision {between different-sized} clouds (Habe \& Ohta 1992, Anathpindika 2010).
In this scenario, high-mass stars are formed in the compressed layer created by {the} collision of two clouds.
{These stars then}{ ionize the parent clouds{,} leading to {the formation of} bubble-like H\,\emissiontype{II} regions} (see Figure 12 of Torii et al. 2015). 
{This scenario has also been suggested as} the formation mechanism {for} many mid-infrared bubbles (e.g., N37,  Baug et al. 2016; Sh2-48, Torii et al. 2018b; RCW 166, Ohama et al. 2018a; RCW 79, Ohama et al. 2018b; S116-117-118, Fukui et al. 2018b; N35, Torii et al. 2018a; N49, Dewangan et al. 2017; and N4, Fujita et al. 2018, in preparation).

{Thus} are many different star-formation scenarios {may be related to} the bubbles, {but} the dominant process is not clear.

\subsection{S44 as {an isolated mid-infrared bubble}}
{S44} is {an isolated} mid-infrared bubble located {at} $(l,b)=$(\timeform{334.523D}, \timeform{0.823D}). {It was} cataloged by Churchwell et al. (2006; see their Figure 2c), and {corresponds} to the closed bubble {identified} from the AKARI 9$\mu$m emission (Hanaoka et al. 2018 submitted). Simpson et al. (2012) also identified S44 as MWP1G334525+008255 by visual inspection {as a part of the "Milky Way project" (see also Kendrew et al. 2012)}. 
{Caswell and Haynes (1985) carried out {a} radio-recombination-line survey toward southern H\,\emissiontype{II} regions, deriving the radio recombination line velocity of $-77$ km s$^{-1}$}.
 From Table 4 of Churchwell et al. (2006), the distance of S44 is estimated {to be} {either} $\sim$ 4.6$\pm$0.5 kpc {(the nearer estimate)} or $\sim$ 10.8$\pm$0.5 kpc {(the more-distant estimate), based on this velocity and} the kinematic-distance model of Brand \& Blitz (1993).
{{Churchwell et al. (2006)} suggested that the {nearer} kinematic distance is {the more} correct choice toward {bubbles,} because it is {more} likely to be detected in continuum emission than {at the grater} distance in the Galactic plane.}
Following previous {studies}, {in this paper we adopt} 4.6 kpc as the distance {to} S44{, which {suggests that it is located} in the Norma spiral arm in the Milky Way (Brown et al. 2014).}
{Hattori et al. (2016) estimated} the total infrared luminosity of S44 to be $1.86\times10^5 (10^{5.27}) L_{\odot}$} {by fitting the} spectral energy distribution (SED) {with} polycyclic aromatic hydrocarbons (PAHs), warm-dust, and cold-dust components. 

Figure 1 shows a three-color composite image of the {\it Spitzer} space telescope observations (GLIMPSE: Benjamin et al. 2003, Churchwell et al 2009; MIPSGAL: Carey et al. 2009), where blue, green, and red correspond to the 3.6 $\mu$m, 8 $\mu$m, and 24 $\mu$m emission, respectively. 
The 3.6 $\mu$m emission mainly traces thermal emission from the stars, {while the 8 $\mu$m emission {traces {the} PAH features in the photo-dissociation region (e.g., Draine 2003, Draine \& Li 2007}, Churchwell et al. 2004). }
{The 8 $\mu$m emission} has  {a} ring-like structure{, with} bright emission {at} the southern edge of S44{, and the diameter of the bubble is about 5 pc in the extended 8 $\mu$m emission}.
The 24 $\mu$m emission traces {hot dust grain{s}} heated by high-mass stars in the H\,\emissiontype{II} region {(e.g, Carey et al. 2009){, and it has an arc-like distribution.}
{A} bright infrared source exists at $(l,b)=(\timeform{334.46D}, \timeform{0.88D})$ {on the} western side of S44, which {corresponds to {the} OH/IR star (OH 334.458+0.877) identified {from} the {OH maser} survey {at 1612.231MHz} using the Australia Telescope Compact Array (ATCA; Sevenster et al. 1997b)}.
The relationship {to} this infrared source {to} S44 is not clear.
{We {also} find pillar-like structures in the $8\ \mu$m emissions at the edge of the bubble {that are} elongated {in} the direction of {the} center, with length{s} of $\sim$ 1pc ({The white dotted arrows in Figure 1b}).}

S44 is {an isolated} mid-infrared bubble, {for which} the {3-dimensional} spatial distribution, velocity structures, and physical properties {of} {the} associated molecular gas {have not {yet} been determined}.
In this paper, we carried out new CO observations {in the direction of} S44 {using} the NANTEN2, Mopra, and ASTE radio telescopes.
This paper is organized as follows: Section 2 describes {the} observational information and Section 3 {{presents the} observed cloud properties} and comparison{s} with {observations at} other wavelength{s}. Section 4 discusses {the} formation mechanism of S44 based on a star-formation scenario, and Section 5 concludes this paper.

\section{Observations}
\subsection{NANTEN2 $^{12}$CO $J=$ 1--0 observations}
We carried out $^{12}$CO $J=$ 1--0 (115.271 GHz) observations with the NANTEN2 4 m millimeter/sub-millimeter radio telescope located in Chile {and} operated by Nagoya University. The observations were made {using} the on-the-fly (OTF) mode from May 2012 to December 2012 as a part of {a} Galactic plane survey. 
The half-power beam width (HPBW) is \timeform{2.7'} at 115 GHz. This corresponds to 3.6 pc at the distance of 4.6 kpc. 
The front end was a cooled{, 4 K} superconductor-insulator-superconductor (SIS) mixer receiver. 
The system temperature including the atmosphere was $\sim$ 250 K in the double-side band (DSB) toward the zenith. 
The backend was a digital-Fourier-transform spectrometer (DFS) with 16384 channels{, each with a} 1 GHz bandwidth. The velocity coverage and resolution were $\sim 2600$ km s$^{-1}$ and 0.16 km s$^{-1}$ at 115 GHz, respectively. {We confirmed that} the pointing accuracy was better than \timeform{15"} {from} daily observations toward IRC +10216 and the Sun. 
We used the chopper-wheel method to calibrate the antenna temperature $T_{\rm a}^*$ (Penzias \& Burrus 1973, Ulich \& Haas 1976, Kutner \& Ulich 1981). 
{We calibrated} the absolute intensity fluctuation {using} daily observations of IRAS 16293-2422 [$\alpha_{\rm J2000} = \timeform{16h32m23.3s} , \delta_{\rm J2000} = \timeform{-24d28'39. 2"}$], and {we converted} the intensity scale into {main-beam temperatures} $T_{\rm mb}$ by assuming its peak to be $T_{\rm mb} = $18 K (Ridge et al. 2006). {The {typical intensity uncertainty of the data sets is $\sim$ 20\%} from the NANTEN2 errors and reference data.}
{We smoothed} the data cube with a Gaussian kernel of \timeform{90"}, and the final beam {resolution} {was} \timeform{180"} (FWHM).
The typical root-mean- square (rms) noise level {was} $\sim$ 1.2 K {after} smoothing and velocity-channel binning down to} 0.43 km s$^{-1}$ .
  
\subsection{Mopra $^{12}$CO {and} $^{13}$CO $J=$ 1--0 observations}
{In July 2014, we observed a} \timeform{12'} $\times$ \timeform{12'} area toward S44 using {the} $^{12}$CO and $^{13}$CO $J =$ 1--0 transition{s} (115.271 {and} 110.201 GHz) {using the Mopra 22 m telescope in the OTF mapping mode at }{Australia Telescope National Facility (ATNF)} located in {the Warrumbungle Mountains.
{The HPBW {was} $\sim$ 33$\pm$ \timeform{2"} at 115 GHz{, as} {measured} from planetary observations (Ladd et al. 2005).}  This corresponds to 0.7 pc at the distance of 4.6 kpc. 
The front end was a high-electron-mobility-transistor (HEMT) receiver covering {the} 3-mm band. 
The typical system noise temperature $T_{\rm sys}$ was $\sim$ 600 K in the single-side band (SSB).
The back end system of {the} Mopra Spectrometer ({MOPS}) has 4096 channels across 137.5 MHz in each of two polarization{s. The} velocity resolution was 0.088 km s$^{-1}${,} and the velocity coverage was 360 km s$^{-1}$ at 115 GHz.
The {pointing} accuracy was checked to be within \timeform{5"} by observing the SiO 86 GHz (3 mm) maser sources {in} AH Sco [$\alpha_{\rm J2000} =\timeform{17h11m17.16s}, \delta_{\rm J2000} = \timeform{-32D19'30.72"}$].
{We checked} the intensity variation by observing {M17SW [$\alpha_{\rm J2000} =\timeform{18h20m23.1s}, \delta_{\rm J2000} = \timeform{-16D11'37.2"}$] and the peak of RCW 79 [$(l,b) = (\timeform{308.760D}, \timeform{0.546D})$].
{Typical data fluctuations in the integrated intensity were about $\sim$ 20\%}. 
{We smoothed} the data cube with a Gaussian kernel of \timeform{30"}, and the final {resolution {was}} \timeform{45"} (FWHM).
We converted {the intensity} from antenna temperatures ($T_{\rm a}^*$) {to} main-beam temperature{s} ($T_{\rm mb}$) by {applying {the} relation} $T_{\rm mb}=T_a^*/ \eta_{\rm mb}$, {assuming} a main-beam efficiency($\eta_{\rm mb}$) of 0.42 for 115 GHz (Ladd et al. 2005).
The rms noise level was $\sim$ 0.76 K for $^{12}$CO $J =$ 1--0 and $\sim$ 0.53 K for $^{13}$CO $J =$ 1--0, with a velocity resolution of 0.43 km s$^{-1}$.

\subsection{ASTE $^{12}$CO $J=$ 3--2 observations}
{In June 2014,} we performed $^{12}$CO $J=$ 3--2 (345.796 GHz) observations {toward an area of \timeform{10'}$\times$ \timeform{10'} {around} S44 using} the Atacama Sub-millimeter Telescope Experiment (ASTE, Ezawa et al. 2004, 2008, Kohno et al. 2004) located in Chile. 
{The HPBW} {was} $\sim$ \timeform{22"} at 345 GHz.  This corresponds to 0.5 pc at the distance of 4.6 kpc.
The front end was the two-side band (2SB) SIS mixer called "CATS 345" (cartridge-type-sideband-separating receiver at 345 GHz, Inoue et al. 2008). 
The typical system temperature was {300 K at 345 GHz} in the single-side band (SSB) {mode}. The back end was the XF-type digital spectro-correlator, “MAC”, with 1024 channels {each} of 128 MHz {width.} The velocity resolution was 0.11 km s$^{-1}$, and the velocity coverage was 111 km s$^{-1}$ at 345 GHz (Sorai et al. 2000). {We checked} the {pointing} accuracy to be within \timeform{2"} by observing RAFGL 4202 [$\alpha_{\rm J2000} =\timeform{14h52m23.82s}, \delta_{\rm J2000} = \timeform{-62D04'19.2"}$]. 
{We checked} the intensity {calibration} by observing {W44 [$\alpha_{\rm J2000} =\timeform{18h50m46.1s}, \delta_{\rm J2000} = \timeform{01D11'11.0"}$]. {Typical data fluctuations in {the peak intensity} were about $\sim \ 10\%$}.
{We convolved} the intensity scale into $T_{\rm mb}$ by assuming the W44 peak to be $T_{\rm mb} = $35.5 K (Wang et al. 1994). 
The rms noise level was $\sim$ 0.3 K for $^{12}$CO $J =$ 3--2, with a velocity resolution of 0.43 km s$^{-1}$.   

\subsection{{Archived} data sets}
{We used the following {archived} datasets to compare with the CO data{,} i.e., {to} the near- and mid-infrared data from {the} {\it Spitzer} space telescope {(GLIMPSE {at} 3.6 $\mu$m, and 8.0 $\mu$m, Benjamin et al. 2003, Churchwell et al. 2009; MIPSGAL {at} 24 $\mu$m, Carey et al. 2009)}. {We  obtained} the 870 $\mu$m radio-continuum-emission data with the Atacama Pathfinder Experiment (APEX) Telescope Large Area Survey of the GALaxy {(ATLASGAL, Schuller et al. 2009)}, and the 843 MHz (36 cm)
 radio continuum emission data from {the Sydney University Molonglo Sky Survey (SUMSS: {Bock et al. 1999}) observed with the Molonglo Observatory Synthesis Telescope (MOST). We summarize the observational properties and archiv{al} information in Table 1.}}

\section{Results}
\subsection{{CO distributions and velocity structures toward S44}}
{Figure 2a shows a large scale $^{12}$CO $J=$1--0 integrated intensity map obtained by NANTEN2. The CO cloud {is distributed over about 50 pc} from the northern to southern side of the bubble {with} a peak at $(l,b)=(\timeform{334.52D}, \timeform{0.77D})$. }
Figure 2b presents a detailed CO distribution {obtained} with Mopra. The cloud distribution delineates the 8 $\mu$m emission at the southern side of the bubble, {and the} intensity {is depressed} inside the bubble.
In order to investigate {{the detailed morphologies} of the parent clouds, we focused on the molecular {gas} around the bubble using the high-spatial-resolution data sets obtained by Mopra.} {Figures 3 and 4} {show} velocity-channel maps of $^{12}$CO and $^{13}$CO $J=$1--0, respectively. 
The cloud distribution {outlines} the bubble shape of the 8 $\mu$m emission, and the southern part {(Figures 3d, 3e, 4d, and 4e)} is more intense than the northern part of the bubble ({Figures 3b, 3c, 4b, and 4c}). The $^{13}$CO emission {coincides with the most} intense part{s} of the $^{12}$CO $J=$ 1--0 emission. We {obtain} the velocity difference between the northern and southern part{s} of the bubble from the velocity range of $-86.4$ {to} ${-77.8}$ km s$^{-1}$ ({Figures 3b-e and 4b-e}).
{Figure 5 shows the first-moment map of $^{13}$CO $J=$ 1--0 and {several} spectra.}
The $^{13}$CO {map} is useful {for investigating} velocity gradient{s} because it delineates the denser regions of the parent clouds. 
{From the first-moment map (Figure 5a) and the spectra (Figure 5b,5c, and 5d),} we can {see} that two velocity components exist around the bubble. 
Focusing on the eastern side of the bubble, we can clearly {identify both} velocity components {in} the same region (Figure 5c).
{We therefore} suggest that these velocity difference{s do not represent} a velocity gradient of a single cloud but {instead are} two independent components around the bubble. {Because of their velocities, we hereafter designate} these two clouds as the “$-84$ km s$^{-1}$ cloud" and {the} “$-79$ km s$^{-1}$ cloud", respectively, {and they} are most likely to be associated with the bubble.

{{Figure 6a} {shows the two $^{13}$CO $J=$ 1--0 clouds} superposed on the Spitzer 8 $\mu$m image.
{{The $-84$ km s$^{-1}$ and $-79$ km s$^{-1}$} {clouds} overlap on the eastern side of the bubble.}
{Figures 6b and 6c} present position-velocity diagrams {for which the} integration ranges focus on the overlapping part{s} of the two clouds. 
{The velocity of {the radio recombination line ($-77$ km s$^{-1}$ from Caswell and Haynes 1987}) is also shown in the position-velocity diagram ({Figures 6b and 6c}).
{We note the cavity-like structure around the bubble, and we find that {the} two clouds {are connected} to each other {by a} bridging feature at intermediate velocities.} The cavity-like structures ({a few km s$^{-1}$}) around the {radio-recombination-line} velocity may be caused by ionization from the exciting star(s), and the bridging feature indicates that these two clouds may be interacting with each other around the bubble.}


\subsection{{Physical properties of {the} two clouds}}
We calculated the physical properties of the two molecular clouds using {the} $^{12}$CO {and} $^{13}$CO lines assuming local {thermodynamical} equilibrium (LTE). {We used} the following procedures  (e.g., Wilson et al. 2009) {to derive} them above the $3\sigma$ noise level.
{First, we obtained the} excitation temperature $T_{\rm ex}$ from the $^{12}$CO peak intensity $T_{\rm mb}(\rm ^{12}CO peak)$, assuming {that} the $^{12}$ CO $J=$ 1--0 line is optically thick:
\begin{eqnarray}
T_{\rm ex} &=& 5.5 \bigg/ \ln \left(1+ {5.5 \over T_{\rm mb}(\rm ^{12}CO peak) + 0.82 }\right) {\    [\rm K]}.
\label{eq:75}
\end{eqnarray}
The excitation temperature{s} of {the $-84$ km s$^{-1}$ and $-79$ km s$^{-1}$ {clouds} are estimated {to be} 8 - 13 K and 8- 25 K, respectively.}
The optical depth{s $\tau_{13}$} of the $^{13}$CO emission at {each} velocity channel {are} calculated {from the} following equation {for} {the} $^{13}$CO brightness temperature [$T_{\rm mb}(v)$]:
\begin{eqnarray}
\tau_{13} (v) &=& -\ln \left[1-{T_{\rm mb}(v) \over 5.3} \left\{ {1 \over \exp({5.3 \over T_{\rm ex}})-1}-0.16 \right\}^{-1} \right].
\label{eq:75}
\end{eqnarray}
{We then} calculated the $^{13}$CO column density {$N ({\rm ^{13}CO})$ for all the velocity} channels {{for which the} resolution $\Delta v$ is} 0.43 km s$^{-1}$:
\begin{eqnarray}
N ({\rm ^{13}CO}) &=& 2.4 \times 10^{14} \times \sum_v {T_{{\rm ex}} \tau_{13} (v) \Delta v \over 1-\exp \left(-{5.3 \over T_{\rm ex}} \right) } {\   [\rm cm^{-2}]}{.}
\label{eq:75}
\end{eqnarray}
{We converted} {$N ({\rm ^{13}CO})$} into H$_2$ column {densities} $N(\rm H_2)$ assuming the {CO abundance {ratio to be} [$^{12}$CO]/[H$_2$] $= 10^{-4}$ (e.g., Frerking et al. 1982; Pineda et al. 2010) and the {isotope} abundance ratio {to be} $[^{12}$C]/[$^{13}$C$] = 77$ (Wilson \& Rood 1994). 
{We find the} peak column {densities} of {the $-84$ km s$^{-1}$ and $-79$ km s$^{-1}$ cloud{s} {to be} $1 \times 10^{22}$ cm$^{-2}$ and $5 \times 10^{22}$ cm$^{-2}$, respectively.}
Finally, we estimated the mass{es} of {the} two cloud{s} using {the} following equation:
\begin{eqnarray}
M =  \mu_{\rm H_2} m_{\rm H} D^2 \Omega \sum N(\rm H_2),
\label{eq:75}
\end{eqnarray}
where $\mu_{\rm H_2}=2.8$ is the mean molecular weight {of molecular hydrogen}, $m_{\rm H}=1.67\times 10^{-24}$ g is the proton mass, $D=4.6 $ kpc is {the} adopted {distance}, and $\Omega$ is the solid angle {subtended by} the cloud.
{The molecular masses of the $-84$ km s$^{-1}$ and $-79$ km s$^{-1}$ {clouds} are {thus} estimated {to be} $4\times 10^3\ M_{\odot}$ and $3\times 10^4\ M_{\odot}$, respectively. }
{We also calculated the total masses and column densities} from the $^{12}$CO $J=$ 1--0 integrated intensity, assuming the {conversion factor} {to be} {$N_{\rm H_2}/W({\rm CO}) = 2 \times 10^{20}$ $($K km s$^{-1})^{-1}$ cm$^{-2}$, with $\pm 30$\% uncertainty{, where $W(^{12}$CO$)$ is the integrated intensity of the $^{12}$CO $J=$1--0 line} (Bolatto et al. 2013).
The cloud masses estimated from {the} $^{12}$CO and $^{13}$CO emissions {differ by a} factor {of} 3 for the $-84$ km s$^{-1}$ cloud.
This may be an effect of the low-density gas traced by $^{12}$CO $J=$1--0.}
We summarize the physical properties of the two clouds {in} Table 2. 

\subsection{{$^{12}$CO $J=$3--2 distributions and {$^{12}$CO $J=$3--2/1--0 intensity ratios}}}
{{Figure 7} {shows} the velocity-channel map {for} $^{12}$CO $J=$3--2 obtained {with} ASTE. 
The $^{12}$CO $J=$ 3--2 distribution {more clearly} shows  {the} ring features associated with the 8 $\mu$m emission {from} the bubble.
We also find two clumps, at $(l,b)=(\timeform{334.495D}, \timeform{0.835D})$ and $(l,b)=(\timeform{334.505D}, \timeform{0.845D})$, at the western side of the bubble {{in} the velocity range of $-84.2$ {to} $-82.1$ km s$^{-1}$ ({Figure 7c})}, and {a} clumpy structure at $(l,b)=(\timeform{334.53D}, \timeform{0.80D})$ in {Figure 7d}.}
{Figures 8a, and 8b} show the $^{12}$CO $J=$3--2/$^{12}$CO $J=$1--0 ($R_{3-2/1-0}$) intensity ratio maps {for} (a) the {$-84$ km s$^{-1}$ and (b) the $-79$ km s$^{-1}$ cloud}, respectively.
We convolved {the $^{12}$CO $J =$ 3--2 {map} with {a} Gaussian kernel of \timeform{45"}, which is the final beam size of the $^{12}$CO $J =$ 1--0 Mopra data. }
{The $5\sigma$ ({$\sim1.5$ K km s$^{-1}$}) clipping level{ we adopted is shown by the white dotted contour of $^{12}$CO $J=$3--2}.}
{The intensity ratio between the different rotational transition levels of {the} CO lines provide{s} the excitation condition{s in} the CO gas, and {a} high intensity ratio is a good indicator of {a} physical association with the H\,\emissiontype{II} region.}
{The $-84$ km s$^{-1}$ cloud has high {ratios} ($R_{3-2/1-0}$ $\sim$ 1.0--1.2) at the western {edge}
 $(l,b)=(\timeform{334.490D}, \timeform{0.825D})$ and {the} northeastern edge $(l,b)=(\timeform{334.555D}, \timeform{0.830D})$ of the bubble ({Figure 8a}).}
{The $-79$ km s$^{-1}$ cloud also {has} enhanced $R_{3-2/1-0}$ $\sim$ 1.0--1.2{, but} around the southern and western edge{s} of the bubble. 
The distributions of the intensity ratio $R_{3-2/1-0}$ delineate the 8 $\mu$m ring structure, showing {a} steep increase of temperature {inside} the bubble. These results indicate that the two clouds are likely to be {physically} associated with the bubble.}

\subsection{Comparison with the ionized gas and cold dust emissions}
{Figures 9a, and 9b} show comparison{s} {of} {the $-84$ km s$^{-1}$ and $-79$ km s$^{-1}$ cloud{s, respectively,}} with SUMSS 843 MHz (36 cm) continuum images.
{The continuum image} traces {the} free-free emission from the ionized gas heated by the high-mass stars. 
The 843 MHz intensity is enhanced {at} the southern side of the bubble, having {an arc-like structure.} 
{We note that the ionized gas is not spread uniformly inside the 8 $\mu$m shell structure, which is different from other bubbles (e.g., N10 and N21, Watson et al. 2008).}
{The $-84$ km s$^{-1}$ cloud {is distributed along} the northern edge of the bubble. ({Figure 9a}).
The $-79$ km s$^{-1}$ cloud {surrounds} the ionized gas at the southern side of the bubble ({Figure 9b}).}
{Figures 9c, and 9d} show the comparison{s} of {the $-84$ km s$^{-1}$ and $-79$ km s$^{-1}$ cloud{s, respectively.} with the 870 $\mu$m continuum images {obtained with APEX}.
The 870 $\mu$m continuum image {shows} the distribution of  the thermal emission from the cold dust (Schuller et al. 2009).
The distribution of the cold dust {outlines} the shape of the bubble{, with some peaks {coinciding} with the peaks of the $-84$ km s$^{-1}$ and $-79$ km s$^{-1}$ clouds.}
The $-84$ km s$^{-1}$ cloud {is distributed along} the edge of the cold dust emission at {$(l,b)\sim (\timeform{334.555D}, \timeform{0.825D})$} ({Figure 9c}). 
{The peak in the 870 $\mu$m emission} at $(l,b)=(\timeform{334.52D}, \timeform{0.77})$ {on} the southern side of the bubble {corresponds} to the compact source AGAL G334.521+00.769 cataloged by the ATLASGAL survey (Contreras et al. 2013, Urquhart et al. 2014). {{Figure 9d}} shows that AGAL G334.521+00.769 {is embedded in} {one of {the} peaks} of the $-79$ km s$^{-1}$ cloud. {We note that the peaks of {the} radio continuum, CO, and 870 $\mu$m {emission have ordered} distributions moving to lower latitude{s}.}

\subsection{{Estimation of the} spectral type of the exciting star(s)}
We investigated the spectral type of the exciting star(s) embedded {in} the bubble from the radio continuum flux.
If we assume {the} 843 MHz (36 cm) {emission from the bubble} {to be} optically thin, we can estimate {the number of} Lyman continuum photon{s} $N_{\rm Ly}$ from the 843 MHz radio continuum flux $S_{\rm \nu}$ by using {the} following equation (Rubin 1968c, Mezger et al. 1974):
\begin{eqnarray}
\left[{N_{\rm Ly} \over {\rm s}^{-1}}\right] &\sim& 4.761 \times 10^{48} a(\nu, T_e)^{-1} \left[{\nu \over {\rm GHz}}\right]^{0.1}  \left[{T_e \over {\rm K}}\right]^{-0.45} \left[{S_{\nu} \over {\rm Jy} }\right] \left[{D \over {\rm kpc} }\right]^2,
\label{eq:75}
\end{eqnarray}
{where} $a(\nu, T_e)^{-1} $ is {the} ratio of the optical path length for free-free emission from Oster (1961) and Altenhoff et al. (1960), which {for} most cases {is} $a(\nu, T_e)^{-1}  \sim 1$ (Mezger \&Henderson 1967a). 
The {radio continuum} flux $S_{\rm \nu}$ is estimated by drawing contours at intensities of {0.015 Jy/beam ($\sim 5\sigma$).}
If we assume the electron temperature {to be} $T_e=10^4$ K in the H\,\emissiontype{II} region ({Ward-Thompson \& Whitworth 2011}), the {numbers of photons} is  {$N_{\rm Ly} \sim 10^{48.08}$ s$^{-1}$.
{If we {assume} a single object, this figure {suggests} that {the} exciting star {in} S44 has a spectral type of O8.5-9, which {corresponds to} {an} $18$--$19 M_{\odot}$ star, from the stellar parameter{s for} Galactic O stars (Martins et al. 2005, Table 4).} {In this paper, we assume that the ionizing O-star is embedded around the peak of the radio-continuum image.}

{\subsection{Color-color diagram of the 24 $\mu$m sources}
We {constructed a} color-color diagram to distinguish between YSOs and other {objects} toward the 24 $\mu$m sources cataloged {around this bubble} by Robert \& Heyer (2015). {They obtained a} radius of 7 pc ($=$\timeform{5.25'}) around the geometric center position at $(l,b)=(\timeform{334.523D}, \timeform{0.823D})$.
{Figure 10} shows the result {of the} [3.6]-[5.8] versus [8.0]-[24] diagram toward the 24 $\mu$m sources. 
We adopted the YSO criteria from Muzerolle et al. (2004), {who} carried out this classification toward the embedded star-forming region NGC 7129 in the Milky Way. 
We identified the \#1 source {in S44} as {a} Class II YSO at the southern edge of the bubble ({Figures 11 c, and d}). {Taking account of its error bars, the \#2 source located outside the bubble also is possibly a Class II YSO. {The photometric parameters of these sources (\#1 and \#2) are summarized in Table 3. Many of the other 24 $\mu$m sources are categorized as Class III/stellar objects. We will not argue these Class III/stellar objects in this paper, because it is not clear whether they are physically related to the bubble. }}

\section{Discussion}
}

\subsection{Star formation around the bubble and the origin of {the isolated O star(s)}}
{From previous studies of bubbles, {the C\&C and/or RDI processes have} been discussed {as} {mechanisms for star formation at the edge of a bubble created} by {expanding an H\,\emissiontype{II} region} (e.g., Deharveng et al. 2010, Zavagno et al. 2006, 2007). 
In the case of S44, we {find a cold-dust condensation at} 870 $\mu$m at the southern edge of the bubble (AGAL G334.521+00.769).
{Figure 11a} shows the H$_2$ column density {derived} from $^{13}$CO $J=$1--0{, togather} with the contours of cold-dust emission} from 870 $\mu$m. 
{There is clearly} {good spatial} correspondence between the high molecular column {densities} and {the peaks of} cold-dust emission, {which suggests that {star formation around the bubble is likely to be happenig at} the southern edge}, while such cold and {dense dust condensations} are not detected at the northern side of the bubble.}
These observational results are common to other bubbles (e.g., RCW 120, Deharveng et al. 2009; Figueira et al. 2017; RCW 79, Zavagno et al. 2006, Liu et al. 2017).

{Our observations} show two velocity components associated with the bubble {at the} {northern and southern side{s of} the 8 $\mu$m emission} ({Figure 6a}), {together with a} bridging feature connecting {the} two clouds ({Figures 6b and 6c}). These signatures suggest that {the} two clouds are interacting with each other {at the bridging feature}, and {they} are similar {to the} {properties of} other bubbles {that are formed by cloud-cloud collisions} (RCW 120, Torii et al. 2015; RCW 79, Ohama et al.2018b; N4, S.Fujita et al. 2018 submitted). 
{Numerical simulations of a cloud-cloud collision {reproduce} the broad-{line} bridging feature {at} the interface between the two clouds {in} the position-velocity diagram (Haworth et al. 2015a, 2015b; see also the review by {Haworth et al. 2017}) based on the model from Takahira et al. (2014, 2018) and Shima et al. (2017). 
From synthetic CO observations, they showed that the bridging feature is caused by turbulent motions in the compressed layer between the two colliding clouds.} 

{Stellar} feedback from the {exciting star} {may} be an alternative {explanation for} the bridging feature. 
If expanding {motion{s} dominate {the} kinematics of} the bubble, we {expect to observe} a ring-like velocity distribution in the position-velocity diagram.
However, in the {position}-velocity diagram {toward} the center of the bubble, we {do not} find expanding velocity structures {from the CO data corresponding} to the sound speed ($\sim$ 10 km s$^{-1}$) of the ionized gas}({Figure 6c}). {This suggests that {any acceleration caused by stellar} feedback is limited}, which is consistent {with the case of RCW 120} (Anderson et al. 2015a, Torii et al. 2015). {We note that S44 is an isolated bubble, because we {do not} find {extended} infrared emission (Figure 1a), {even} though the molecular clouds {are distributed up to 50 pc beyond the northern and} southern side{s of the} bubble (Figure 2a).}
Hence, {isolated O star(s)} are unlikely to be formed by stellar feedback from other high-mass stars.
The formation of {massive,} dense cores {from an O star or a} star cluster {require} external shock compression (e.g., Zinncker \& Yorke 2007).
{Some {numerical} magneto-hydrodynamical simulations} show that a cloud-cloud collision process {satisfies} {the initial condition{s for} O-star formation} (e.g., Inoue \& Fukui 2013, Wu et al. 2015, 2017a, 2017b).
 {We therefore} {hypothesize} that the two clouds collided with each other and {that} the collision triggered the formation of {an} isolated, {massive} exciting star.}

\subsection{{A cloud-cloud collision scenario}}
{Based on our observational results}, in this section we propose {a scenario in which star formation is triggered by} a cloud-cloud collision.
{From} the similar mid-infrared bubble RCW 120 (Torii et al. 2015){, and} based on the numerical simulation of Habe \& Ohta (1992), our proposed scenario is as follows (see our schematic picture {in} {Figure 12}):
\begin{itemize}
\item {The $-84$ km s$^{-1}$ diffuse cloud, enclosing a dense core, and the $-79$ km s$^{-1}$ cloud approach each other ({Figure 12, stage I}).}
\item  The two clouds collide {with} each other{, creating} a compressed layer in the dense part of the $-84$ km s$^{-1}$ cloud at the interface {between} the two clouds and {forming} a cavity {in} the $-79$ km s$^{-1}$ cloud. The two {clouds} {mix at} the boundary {and form the intermediate velocity component} ({Figure 12, stage II}).
\item A High-mass star are formed in the compressed layer at the interface {between} the two colliding clouds. The parent cloud and {the} surrounding interstellar medium are {then} ionized, leading to the formation of {a} bubble-like structure ({Figure 12, stage III}).
\end{itemize}
{{Figure 11b} shows {the hot-dust} {distribution {at} 24 $\mu$m} {, together} with contours of {the} cold-dust emission {{at} 870 $\mu$m}. 
The $24\ \mu$m {hot-dust emission} has {an} asymmetric distribution at the southern side of the bubble, {that} is more intense than {that at the} northern side. 
We also find {a class II YSO at 24 $\mu$m} (red arrows Figures 11c and d) {that is} {embedded} in the cold-dust condensation {producing the} {870} $\mu$m emission. {This} is similar to the distribution of YSOs emitting {at} 24 $\mu$m embedded in “condensation 1" at the edge of RCW 120 (Deharveng et al. 2009, Figure 10). 
{We note that it is not clear whether a cloud-cloud collision caused the formation of this class II YSO, because it may be a pre-existing star.}

{Torii et al. (2015) and Ohama et al. (2018b) showed that  {remnants} of the {colliding} cloud{s} exist outside the opening{s} of the bubble{s in} RCW 120 and RCW 79.
In the case of S44, the opening (broken) {bubble {seen} in} 8 $\mu$m emission {is not} clearly {comparable to}  RCW 120 {or} RCW 79. 
{We suggest that this difference between {the} closed and broken 8 $\mu$m emission {bubbles} can {be explained by} the projection effect toward the bubble. {Based on {this} assumption, we adopt} the projection angle toward S44 as {\timeform{45D}} in section 4.3 {below}.}

\subsection{{The timescale {for} star formation}}
If we assume {an} inclination angle {of \timeform{45D}, the collisional timescale is about 20 pc$/ (5 \times \sqrt{2}$) km s$^{-1} \sim$ 3 Myr} from the extended cloud size and velocity difference.
On the other hand, if we {assume} the mass-accretion rate of high-mass stars {to be} $2\times10^{-4}\ M_{\odot}{\rm yr}^{-1}$ from the numerical simulation of Inoue et al. (2018), the timescale {for} {high-mass} {star} formation {in} S44 is {($18$--$19 M_{\odot}$)}$/2\times10^{-4}\ M_{\odot}{\rm yr}^{-1}\ \sim\ 0.1$ Myr.
We {thus} suggest that the event of a cloud-cloud collision happen{s on} a long time scale ($\sim$ a few Myr), because the small cloud is decelerate{d} {by conserving the momentum through the collision}, whereas {O-star} formation {has} a {short} timescale {of} $\sim$0.1 Myr.
This is similar to the case of the super {star} cluster NGC 3603, except for the number of O stars and {the} high {H$_2$} column density (Kudryavtseva et al. 2012, Fukui et al. 2014). We propose that S44 {may} be a miniature {version} of {a} super star cluster.






\section{Conclusions}
We summarize the conclusions of the present study as follows:
\begin{enumerate}
\item We made new CO observations toward the mid-infrared bubble S44 using NANTEN2, Mopra, and ASTE. {We identified} two clouds, {at} $-84$ km s$^{-1}$ and $-79$ km s$^{-1}$, {in the direction of} the bubble.
\item The $-84$ km s$^{-1}$ {cloud} shows diffuse CO emission {that} extend{s} outside of the bubble, {with} $R_{\rm 3-2/1-0}$ {greater} than 0.6 {on the} northern side of the bubble.
{From the Mopra and ASTE data sets,} the $-79$ km s$^{-1}$ cloud correspond{s morphologically to} the $8\ \mu$m emission, {with} $R_{\rm 3-2/1-0}$ {greater} than 0.8 around the bubble. The ionized-gas and cold-dust images {exhibit} a spatial correlation {with} the bubble. 
\item We estimate the spectral type of the exciting star {to be} {O8.5-9} ($\sim 20 M_{\odot}$), from the SUMSS 843 MHz (36 cm) radio-continuum flux, if we assume a single object. 
\item The two clouds are {connected} {by a} bridging feature at intermediate {velocities that} overlap on the eastern side of the bubble. These observational signatures are interpreted {as being due to the} {interaction} {between the two clouds}.
\item We {hypothesize} that the two clouds {collided with} each other {3 Myr ago,} triggering {the formation of {the O star(s)} and the} isolated bubble.
A cloud-cloud collision scenario can explain the morphology of the two clouds {and the origin of the isolated O-star).}
\end{enumerate}

\section*{Acknowledgements}
We are grateful to the referee, Dr. Christer Watson, for thoughtful comments on the paper. We are also grateful to Mr. Kazuki Okawa for a useful discussion.
NANTEN2 is an international collaboration of ten universities: Nagoya University, Osaka Prefecture University, University of Cologne, University of Bonn, Seoul National University, University of Chile, University of New South Wales, Macquarie University, University of Sydney, and Zurich Technical University.
The work is financially supported by a Grant-in-Aid for Scientific Research (KAKENHI, No. 15K17607, 15H05694) from MEXT (the Ministry of Education, Culture, Sports, Science and Technology of Japan) and JSPS (Japan Soxiety for the Promotion of Science).
The ASTE telescope is operated by National Astronomical Observatory of Japan (NAOJ).
This work is based on observations made with the Spitzer Space Telescope, which is operated by the Jet Propulsion Laboratory, California Institute of Technology under a contract with NASA. Support for this work was provided by NASA.
The Mopra radio telescope is part of the Australia Telescope National Facility which is funded by the Australian Government for operation as a National Facility managed by CSIRO. The University of New South Wales Digital Filter Bank used for the observations with the Mopra Telescope was provided with support from the Australian Research Council. The ATLASGAL project is a collaboration between the Max-Planck-Gesellschaft, the European Southern Observatory (ESO) and the Universidad de Chile. It includes projects E-181.C-0885, E-078.F-9040(A), M-079.C-9501(A), M-081.C-9501(A) plus Chilean data.
{The authors would like to thank Enago (www.enago.jp) for the English language review.}



\begin{table*}
\tbl{Observational properties of data sets.}{
\begin{tabular}{cccccccccc}
\hline
\multicolumn{1}{c}{Telescope} & Line & HPBW   &  Velocity & RMS noise$^{(2)}$  \\
& &  & Resolution & level \\
\hline
NANTEN2 &$^{12}$CO $J=$ 1--0 \footnotemark[]  &  \timeform{160"}  & 0.16 km s$^{-1}$ & $\sim 1.2$ K  \\
Mopra  &$^{12}$CO $J=$ 1--0\footnotemark[] &  $33\pm$\timeform{2"}$^{(1)}$  &  0.088 km s$^{-1}$& $\sim 0.76$ K   \\
&$^{13}$CO $J=$ 1--0\footnotemark[] & $33\pm$\timeform{2"}$^{(1)}$  &  0.092 km s$^{-1}$& $\sim 0.53$ K \\
ASTE &$^{12}$CO $J=$ 3--2\footnotemark[] &  \timeform{22"}   &  0.11 km s$^{-1}$& $\sim 0.3$ K \\
\hline
\hline
Telescope/Survey & Band  &  Resolution & References &  \\
\hline
{\it Spitzer}/GLIMPSE   & 3.6 $\mu$m & $\sim$\timeform{2"} & [1,2]  \\
{\it Spitzer}/GLIMPSE   &  8.0 $\mu$m  &$\sim$\timeform{2"} & [1,2]  \\
{\it Spitzer}/MIPSGAL   & 24 $\mu$m  & \timeform{6"} &  [3]  \\
APEX/ATLASGAL & 870 $\mu$m  & \timeform{19"} & [4] \\
MOST/SUMSS & 843 MHz  & $\sim$ \timeform{60"} & [5] \\
\hline
\end{tabular}}\label{tab:first}
\begin{tabnote}
\footnotemark[1] Reference : Ladd et al. (2005). \\
\footnotemark[2] {The values of the rms noise levels after smoothing the (space and/or velocity) data sets.}\\
References [1] Benjamin et al. (2003), [2] Churchwell et al. (2009), [3] Carey et al. (2009), [4] Schuller et al. (2009), [5] Bock et al. (1999) \\
\end{tabnote}
\end{table*}

\begin{table*}
{
\tbl{Physical properties of the two clouds}{
\begin{tabular}{cccccccccccc}
\hline
\multicolumn{1}{c}{Name} & $T_{\rm ex}$ & $\tau_{13}$ &  $N({\rm H_2})_{\rm peak}$ & $N({\rm H_2})_{\rm mean}$ & $M(^{13}$CO) &  $M(^{12}$CO) \\
& [K] & &[cm$^{-2}$] & [cm$^{-2}$] & [$M_{\odot}$]  & [$M_{\odot}$]\\
\hline
$-84$ km s$^{-1}$ cloud & 9 &  0.36 & $1 \times 10^{22}$  & $9 \times 10^{20}$ & $4 \times 10^3$& {$1 \times 10^4$}\\
$-79$ km s$^{-1}$ cloud & 12 & 0.38 & $5 \times 10^{22}$  &$7 \times 10^{21}$ & $3 \times 10^4$ &{$3 \times 10^4$} \\
\hline
\end{tabular}}\label{tab:first}
\begin{tabnote}
The excitation temperature $T_{\rm ex}$ and mean column density $N({\rm H_2})_{\rm mean}$ are the averaged values above the $3\sigma$ noise level for each cloud.\\
The optical depths $\tau_{13}$ the averaged values from the integrated velocity range above the $3\sigma$ noise level.
\end{tabnote}
}
\end{table*}

{
{
\begin{table*}
{{
\tbl{Photometric data for the YSO candidates from the 24 $\mu$m sources (Robert \& Heyer 2015) 7 pc from the center position of the bubble.}
{
\begin{tabular}{cccccccccccc}
\hline
\multicolumn{1}{c}{Number} & $l$&$b$  & [3.6] &$\sigma_{3.6}$ &  [5.8] &$\sigma_{5.8}$ & [8.0] &$\sigma_{8.0}$& [24] &$\sigma_{24}$\\
& [deg] &[deg]  & [mag] & [mag] &  [mag] & [mag] & [mag] & [mag] & [mag] & [mag]\\
\hline
1	& 334.502	& 0.761  & 9.568 & 0.041 & 8.505 & 0.033  & 7.544	&0.048 & 3.90 &0.06	\\
2	& 334.605	& 0.826 & 10.958 & 0.086  & 10.764 & 0.084 & 10.299 & 0.055& 6.30 & 0.03	\\
\hline
\end{tabular}}\label{tab:first}
\begin{tabnote}
\end{tabnote}
}}
\end{table*}}}

\clearpage
\begin{figure*}[h]
\begin{center}  \includegraphics[width=18cm]{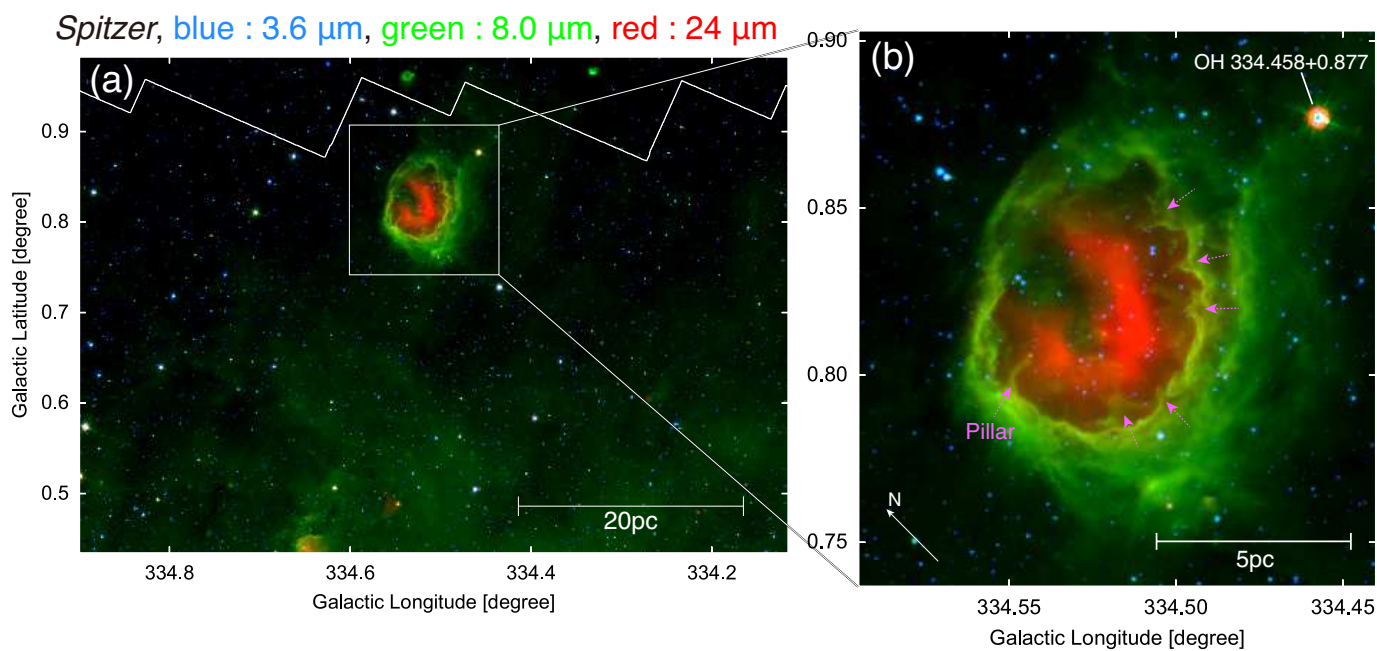}
\end{center}
\caption{(a) Large-scale three-color composite images of S44. Blue, green, and red show the {\it Spitzer}/IRAC 3.6-$\mu$m, {\it Spitzer}/IRAC 8-$\mu$m, and {\it Spitzer}/MIPS 24-$\mu$m results. The {jagged} white line along the Galactic latitude at $b\sim \timeform{0.95D}$ shows the observing {limit} of {\it Spitzer}/MIPS 24-$\mu$m. (b) A close-up image of (a). The colors are the same as in (a). {The pink dotted arrows indicate the pillar-like structures.}}\label{.....}
\end{figure*}

\begin{figure*}[h]
\begin{center} 
\includegraphics[width=18cm]{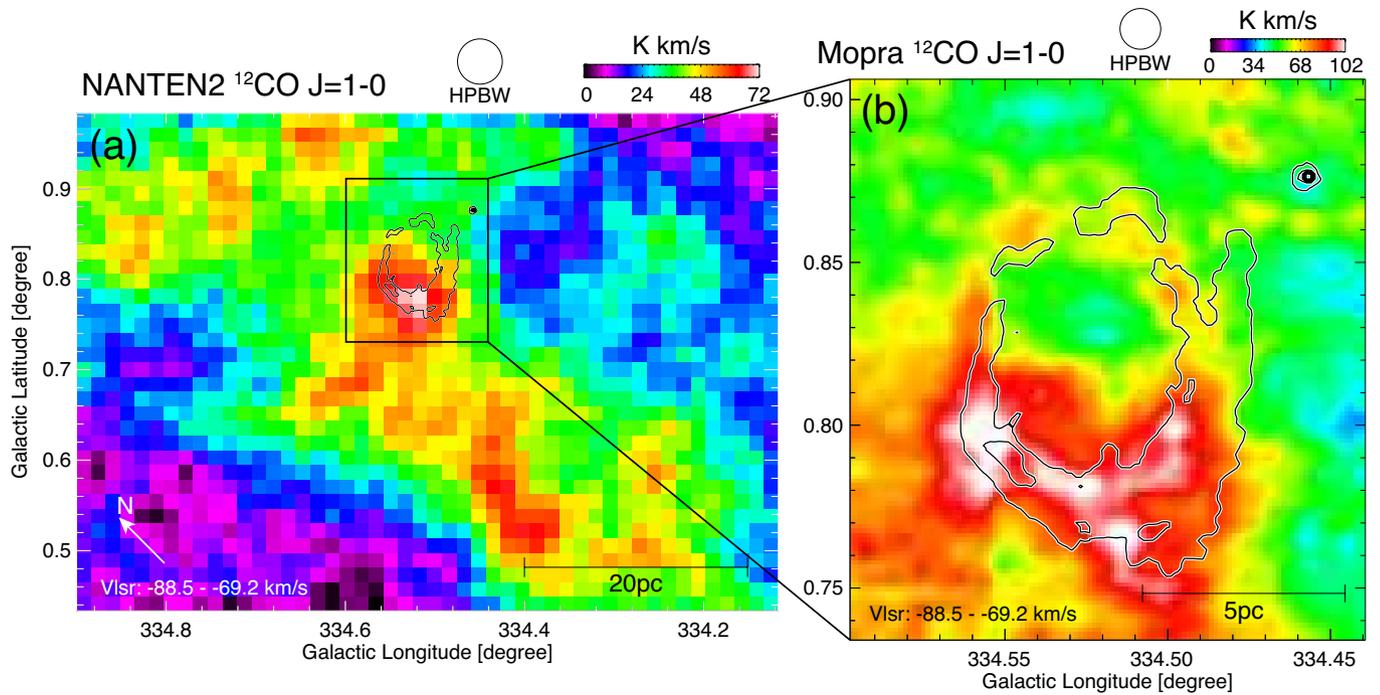}
\end{center}
\caption{(a) Integrated map of the $^{12}$CO $J=$ 1--0 emission {in} the velocity range of $-88.5$ {to} $-69.2$ km s$^{-1}$. The contours show the {\it Spitzer}/IRAC 8-$\mu$m result, where the region used for the 8-$\mu$m emission is indicated by the black box. The lowest contour and contour intervals are 70 MJy/beam and 80 MJy/beam, respectively. The final beam size after convolution is \timeform{180"} (b) Integrated intensity map of the $^{12}$CO $J=$ 1--0 emission with Mopra. The final beam size after convolution is \timeform{45"}}\label{.....}
\end{figure*}


\begin{figure*}[h]
 \includegraphics[width=18cm]{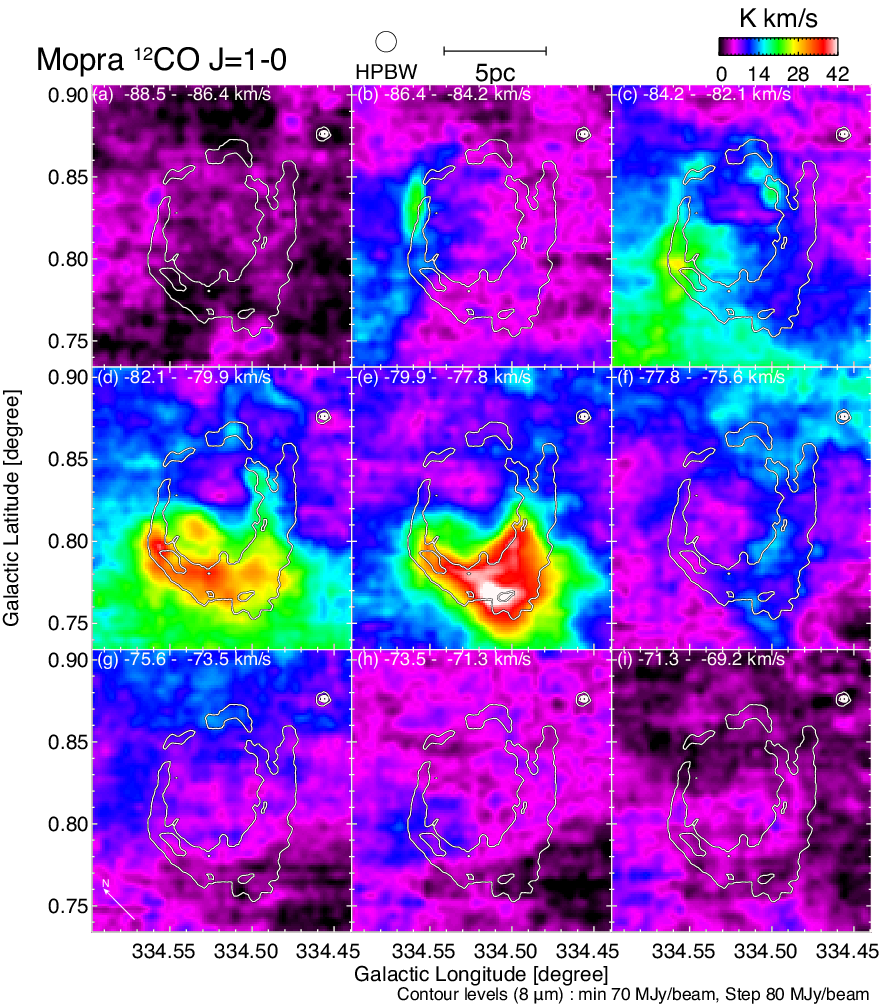}
\caption{Velocity-channel map of the $^{12}$CO $J=$ 1--0 emission, with a velocity step of  2.15 km s$^{-1}$, obtained by Mopra. The contours show the 8$\mu$m emission from {\it Spitzer}/IRAC. The final beam size after convolution is \timeform{45"}. The 1$\sigma$ noise level is $\sim$ 0.7 K km s$^{-1}$ for the velocity interval of 2.15 km s$^{-1}$.}\label{.....}
\end{figure*}

\begin{figure*}[h]
 \includegraphics[width=18cm]{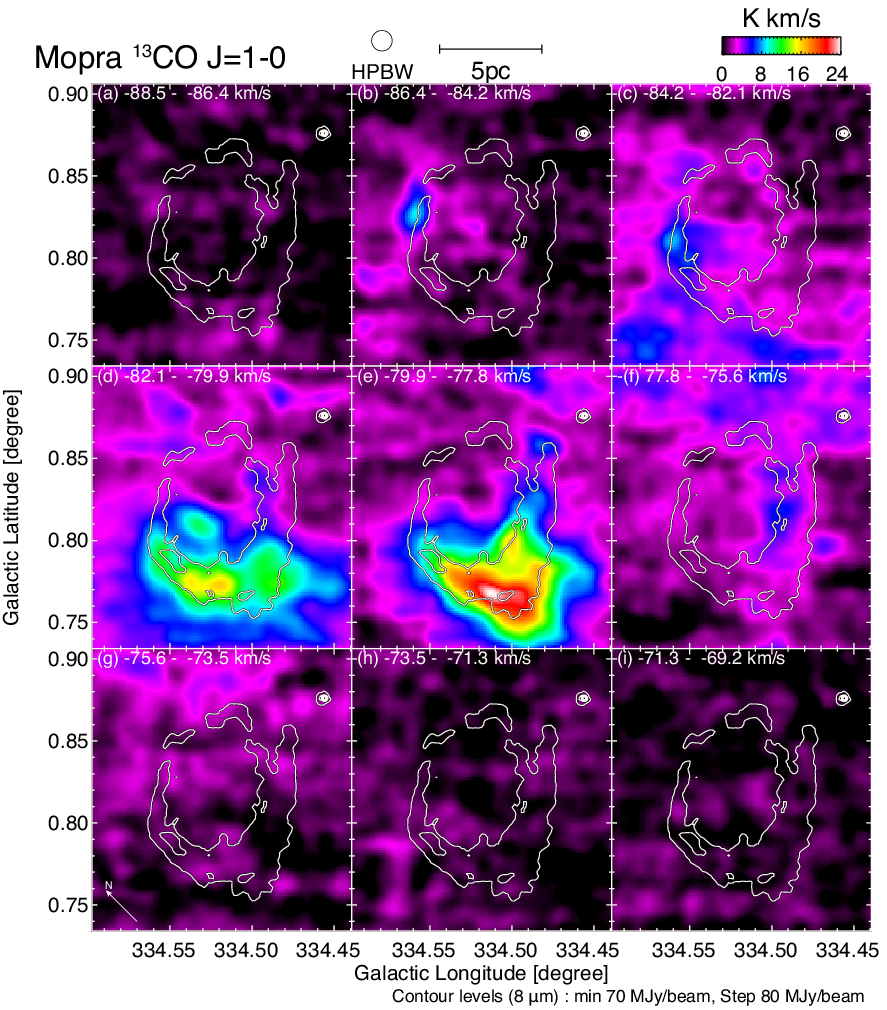}
\caption{Velocity-channel map of the $^{13}$CO $J=$ 1--0 emission, with a velocity step of 2.15 km s$^{-1}$, obtained by Mopra. The contours show the 8$\mu$m emission  {from} {\it Spitzer}/IRAC. The final beam size after convolution is \timeform{45"}. The 1$\sigma$ noise level is $\sim$ 0.5 K km s$^{-1}$ for the velocity interval of 2.15 km s$^{-1}$.}\label{.....}
\end{figure*}


\begin{figure*}[h]
\begin{center} 
 \includegraphics[width=18cm]{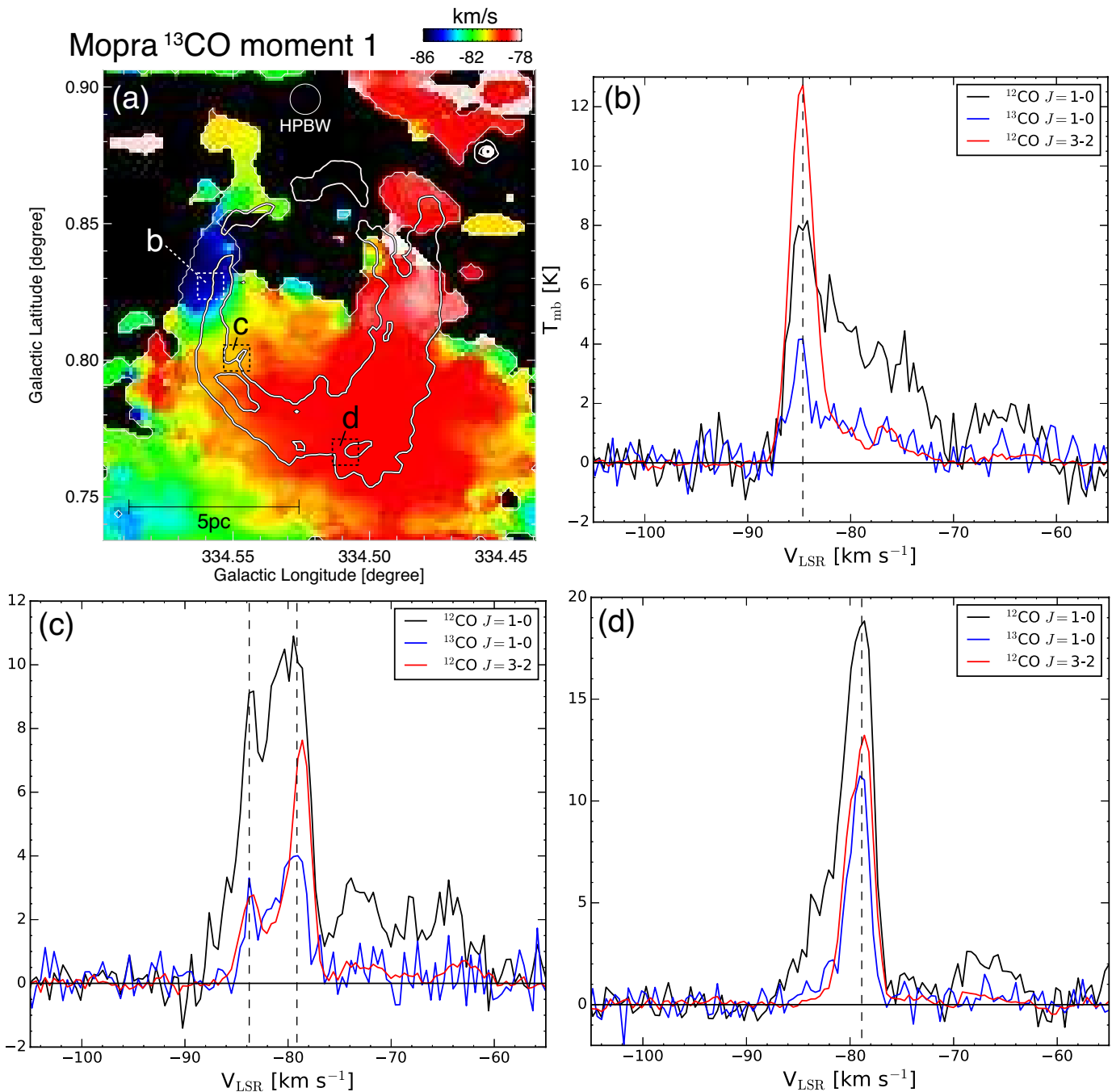}
\end{center}
\caption{(a) The first-moment map of the $^{13}$CO $J =$ 1--0 emission, which {we} created for the velocity range of $-87.64$ to $-77.33$ km s$^{-1}$using the volume voxels with the intensities {greater} than 2.1 K (4$\sigma$). The lowest contour and contour intervals {were} 70 MJy/beam and 80 MJy/beam {for} the {\it Spitzer}/IRAC 8-$\mu$m result. {The boxes show the averaging areas for each profile. (b), (c), and (d) The averaged spectra for $^{12}$CO, $^{13}$CO $J=$1--0, and $^{12}$CO $J=$3--2. The dotted lines indicate the two velocity components at $-84$ km s$^{-1}$ and $-79$ km s$^{-1}$} The size of averaging box is $\timeform{35"} \times \timeform{35"}$.}\label{.....}
\end{figure*}

\begin{figure*}[h]
 \includegraphics[width=18cm]{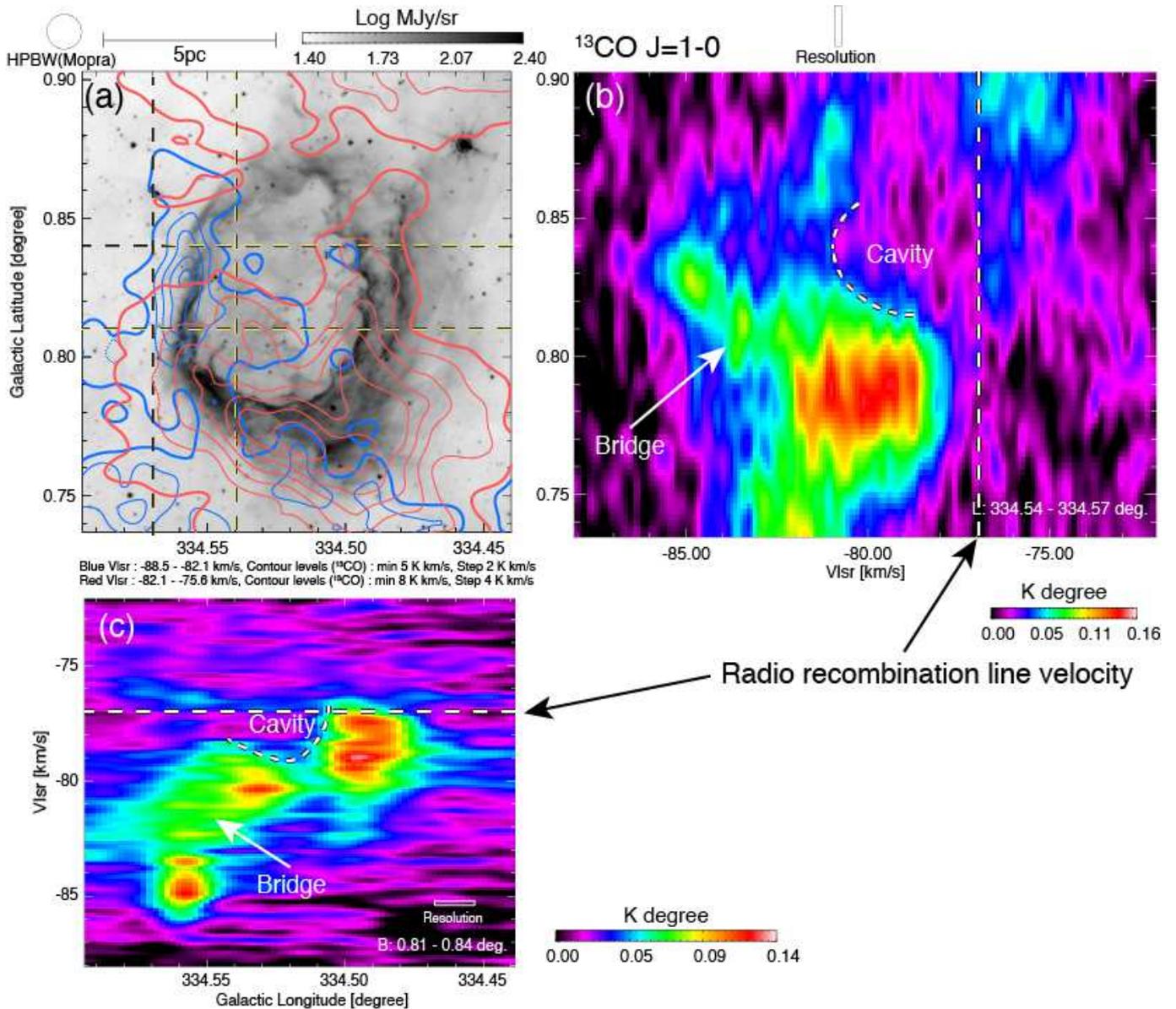}
\caption{(a) Integrated intensity map of $^{13}$CO $J=$ 1--0 obtained by Mopra for the $-84$ km s$^{-1}$ cloud (blue contours) and the $-79$ km s$^{-1}$ cloud (red contours) superposed on the {\it Spitzer} $8 \mu$m emission. The yellow dashed lines show the integration ranges in latitude and longitude. (b) Galactic latitude-velocity diagram integrated over the {longitude} range from \timeform{334.54D} to \timeform{334.57D}. {The $1 \sigma$ noise level is $\sim$ 0.004 K degree for the longitude interval of \timeform{0.03D} .} (c) Galactic longitude-velocity diagram integrated over the latitude range from \timeform{0.81D} to \timeform{0.84D}. The dashed lines represent the {radio- recombination-line velocity ($-77$ km s$^{-1}$) from Caswell and Haynes (1987)}. The spatial and velocity resolution are smoothed to \timeform{52"} and 0.18 km s$^{-1}$, respectively.  {The $1 \sigma$ noise level is $\sim$ 0.004 K degree for the latitude interval of \timeform{0.03D}.}}\label{.....}
\end{figure*}


\begin{figure*}[h]
 \includegraphics[width=18cm]{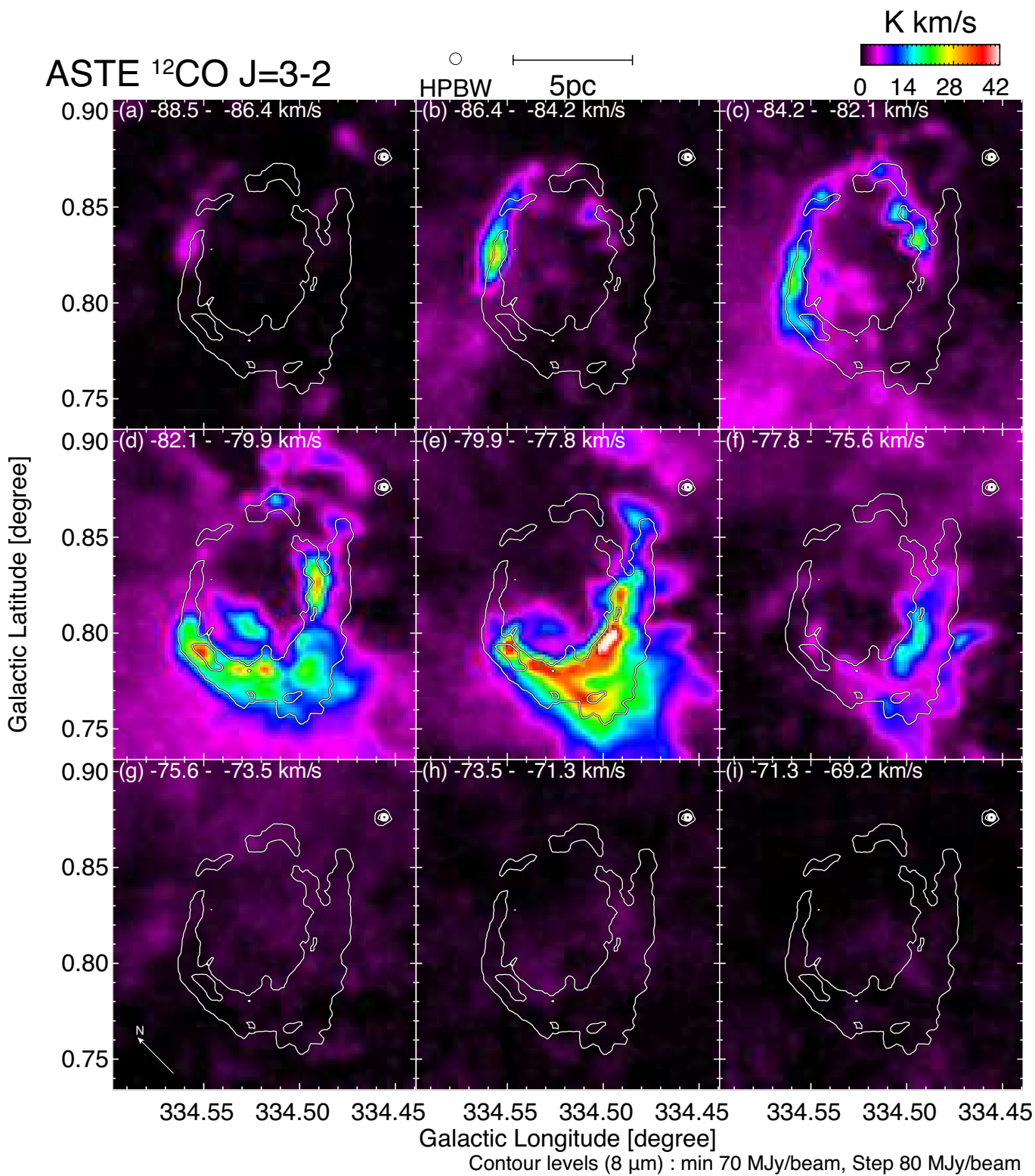}
\caption{Velocity-channel map of the $^{12}$CO $J=$ 3--2 emission, with a velocity step of  2.15 km s$^{-1}$, obtained by ASTE. The contours show {the} 8 $\mu$m emission {from} {\it Spitzer}/IRAC. The final beam size is \timeform{22"}. The 1$\sigma$ noise levels are $\sim$ 0.3 K km s$^{-1}$ for the velocity interval of 2.15 km s$^{-1}$.}\label{.....}
\end{figure*}

\begin{figure*}[h]
 \includegraphics[width=16cm]{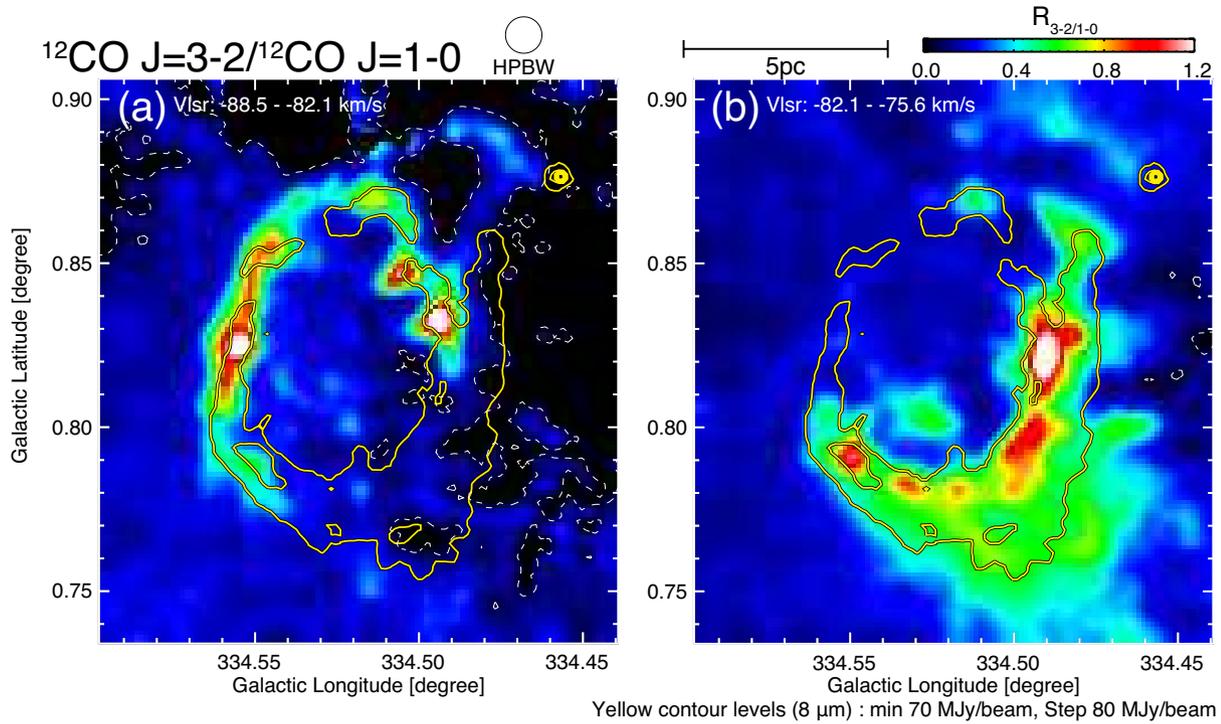}
\caption{(a),(b) Intensity ratio map of $^{12}$CO $J =$ 3-2/$^{12}$CO $J =$ 1--0 {from} ASTE and Mopra for the $-84$ km s$^{-1}$ cloud (a) and the $-79$ km s$^{-1}$ cloud (b). The final beam size after convolution is $\sim$ \timeform{45"}. {The 5$\sigma$ ($\sim 1.5$ K km s$^{-1}$) clipping levels adopted are shown by the dotted white contour of $^{12}$CO $J=$3--2.} {The lowest yellow contour and intervals} are 70 MJy/beam and 80 MJy/beam for the {\it Spitzer}/IRAC 8-$\mu$m result.}\label{.....}
\end{figure*}

\begin{figure*}[h]
 \includegraphics[width=16cm]{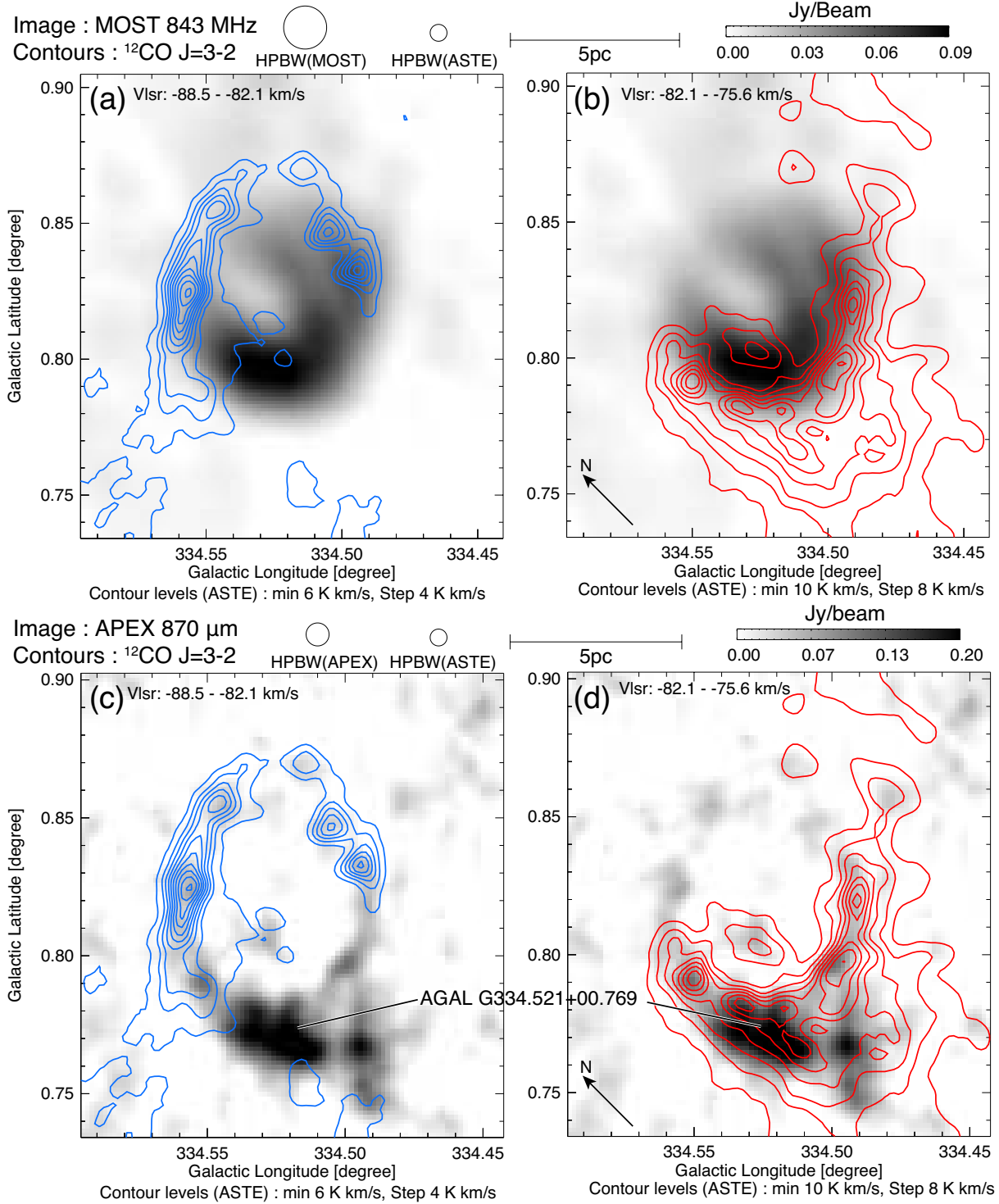}
\caption{(a), (b) Integrated intensity map of $^{12}$CO $J=$ 3--2 (contours) obtained with ASTE superposed on the MOST 843 MHz continuum image. (c), (d) Integrated intensity map of $^{12}$CO $J=$ 3--2 (contours) obtained by ASTE superposed on the APEX 870 $\mu$m continuum image. The blue and red contours represent the $-84$ km s$^{-1}$ cloud and the $-79$ km s$^{-1}$ cloud, respectively.}\label{.....}
\end{figure*}

\begin{figure*}[h]
 \includegraphics[width=13cm]{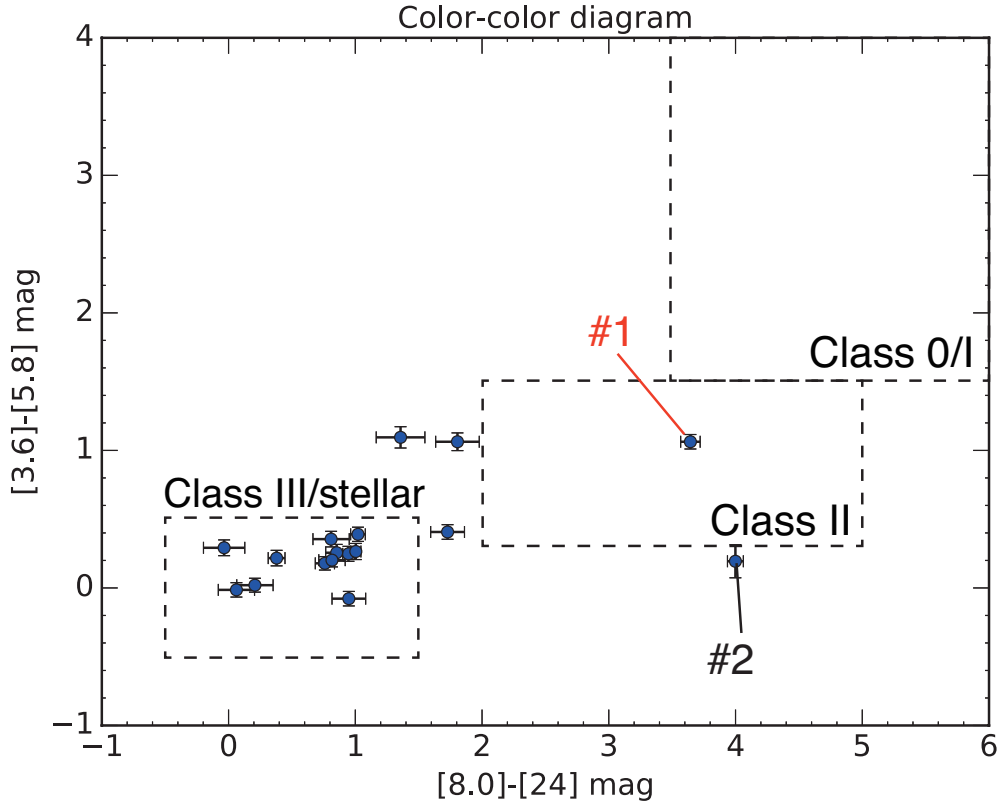}
\caption{{The color-color diagram ([3.6]-[5.8] versus [8.0]-[24]) for the 24 $\mu$m sources cataloged by Robert \& Heyer (2015) around the bubble. The dotted boxes show the classification of YSOs from Muzerolle et al. (2004).}}\label{.....}
\end{figure*}

\begin{figure*}[h]
 \includegraphics[width=18cm]{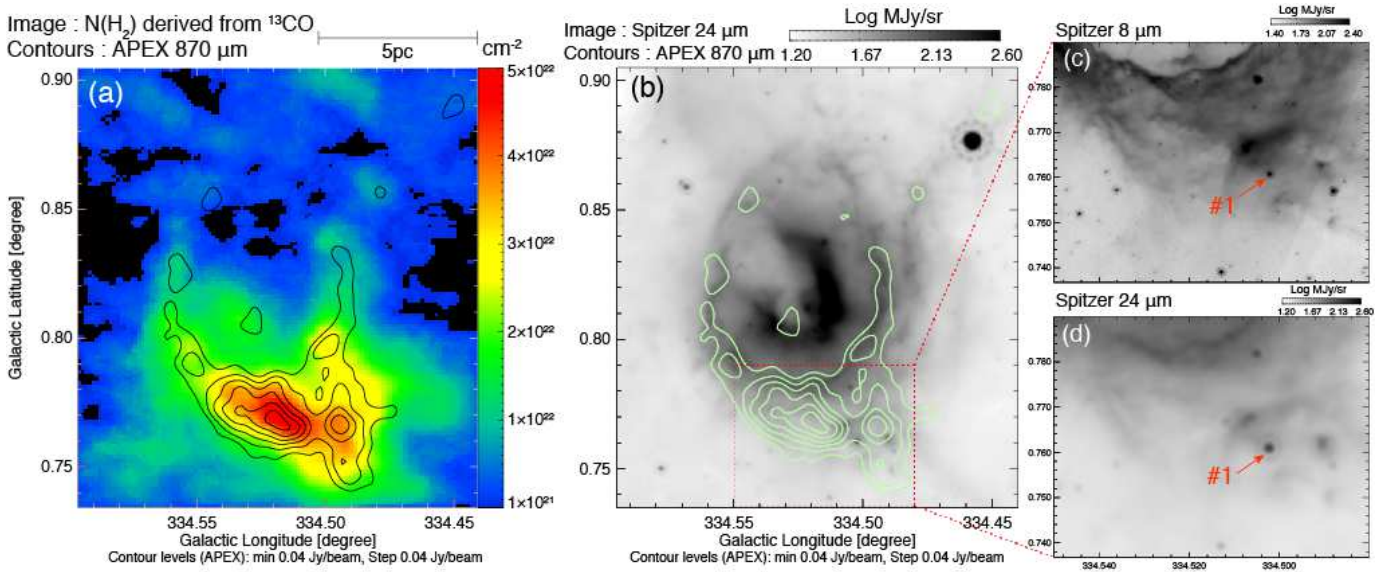}
\caption{(a) The APEX 870 $\mu$m image (contours) superposed on the H$_2$ column density image from $^{13}$CO $J=$1--0 with the velocity range of $-88.5$ to $-69.2$ km s$^{-1}$.  (b) The APEX 870 $\mu$m image (contours) superposed on the {\it Spitzer} $24\ \mu$m emission. The red dotted {square} shows the {sources detected in} the $24\ \mu$m image embedded in the cold dust condensation. {(c) Close-up 8 $\mu$m and (d) 24 $\mu$m images, respectively, at the southern edge of the bubble. {The YSO is indicated by a red arrow in panels (c) and (d).}}}\label{.....}
\end{figure*}

\begin{figure*}[h]
\begin{center} 
\includegraphics[width=18cm]{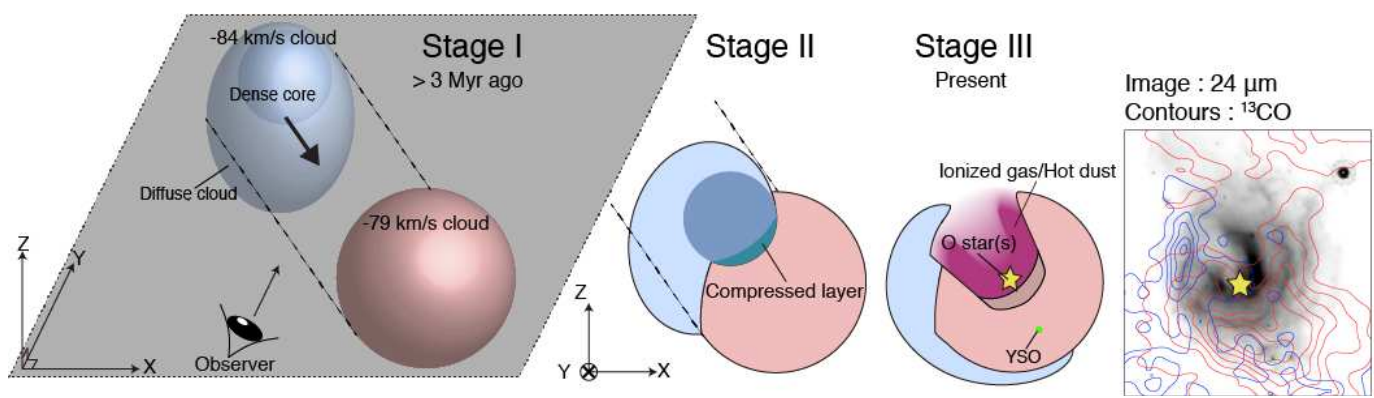}
\end{center}
\caption{Schematic image of a cloud-cloud collision scenario based on Habe \& Ohta (1992) and Torii et al. (2015, 2017a). {Stage I : a 3D image of the initial condition of the two clouds. Stage II : a 2D image of the two clouds in the X-Z plane at the time when the two clouds collide with each other. Stage III : a 2D image of the bubble in the X-Z plane. The Y-axis corresponds to the line of sight.} {The {final} panel shows the observational result for the two clouds superposed on the 24 $\mu$m image.} }\label{.....}
\end{figure*}





\end{document}